\documentstyle[12pt,moriond]{article}
\begin{document}
\heading{THE FAINT END OF THE LUMINOSITY FUNCTION IN CLUSTERS} 

\author{N. Trentham $^{1}$ } {$^{1}$ Institute of Astronomy, Cambridge,
       U.~K.}  {$^{ }$ }

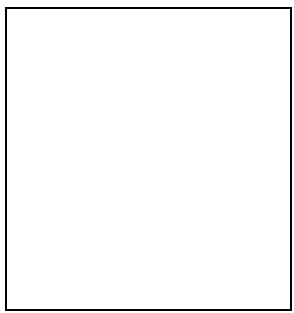
\begin{moriondabstract}
I review recent measurements of the faint end of the galaxy luminosity
function in galaxy clusters.
Evidence is presented that the luminosity function 
of galaxies in the central parts
of clusters is remarkably constant between clusters
and that
this luminosity function is steep at bright
and faint magnitudes and shallow in-between.  The curvature is
highly significant -- neither a power-law nor a Schechter function
is consistent with the
data.  At no magnitude does $\alpha = -1$ fit the data well.
The faintest galaxies in all clusters that have been
studied are dwarf spheroidal galaxies.
\end{moriondabstract}

\section{Introduction -- luminosity functions}
The luminosity function (LF) $\phi (L)$ of galaxies is defined as the
number density of galaxies per unit luminosity $L$: the number of
galaxies in volume ${\rm d}V$ with luminosity between $L$ and $L +
{\rm d}L$ is $\phi(L) {\rm d}L {\rm d}V$.

The LF represents one of the most direct interfaces between studies
of dwarf galaxies and cosmology, the two subjects of this conference.
This statement follows directly from the form of the fluctuation
spectrum -- most plausible theories of structure formation
(e.g.~cold dark matter [3]) predict a fluctuation
spectrum whose power spectrum $\vert \delta (k) \vert^{2} \propto k^n$
has index $n \sim -2$ on scales of galaxies.  This value of $n$
simultaneously requires that low-mass systems be far more numerous
than high-mass ones, so that the statistical properties of low-mass
systems are a powerful probe of galaxy formation theories.

Consequently one of the most fundamental predictions of models of
galaxy formation and evolution is the LF, particularly at faint
magnitudes (e.g.~refs.~6,13,48--50). 
These models
generally assume a power spectrum of primordial dark matter
fluctuations (say, from CDM theory) 
and then adopt prescriptions
for the behavior of the baryonic component (specifically, things like
the effects of gas dissipation, star-formation efficiencies, feedback
from supernovae, and the temperature-density structure of the
intergalactic medium need to be quantified).

Significant differences in the theoretical luminosity functions from
different models are seen, even when similar dark matter power
spectra are assumed, because of differing prescriptions for the
behavior of the baryonic component.  For example Babul \& Ferguson
[2] suggest a faint-end slope of $\alpha \sim -2.6$ at faint 
magnitudes, whereas White \& Kauffmann [49] compute values more
like $\alpha \sim -1$ for their dwarf suppression models (here
$\alpha$ is the logarithmic slope of the LF: $\alpha = 
{\rm d} \log \phi(L) / {\rm d} \log L$).  Therefore measuring $\alpha$
as accurately as possible at faint magnitudes seems like a worthwhile
exercise.

\section{Introduction -- dwarf galaxies}

I will introduce this section of the review by referring to the
absolute magnitude vs.~central surface-brightness plot that at this
conference was referred to as the Binggeli diagram.  As Figure 1 of
Binggeli's 1994 review paper [5, see also 14,15] 
shows, stellar systems generally fall into four
distinct regions on this diagram:
\vskip 1pt
\noindent
{\bf (1) Elliptical galaxies and bulges.~}
These form a sequence of decreasing central surface-brightness with
increasing luminosity.  Absolute magnitudes range from about
$M_B \sim -25$ (e.g.~cD galaxies in the centers of clusters like
NGC 6166 in Abell 2199, which have faint central surface-brightnesses)
to $M_B \sim -15$ (compact low-luminosity ellipticals like M32 which
have bright central surface-brightnesses).
\vskip 1pt
\noindent
{\bf (2) Globular clusters.~}
These have high central surface-brightnesses ($\mu_B \leq 18$ mag
arcsec$^{-2}$), like ellipticals and bulges but have much fainter
absolute magnitudes ($M_B > -10$).  These form a sequence of higher central
surface-brightness with increasing luminosity, which is in the opposite
sense to the correlation for ellipticals and bulges.
\vskip 1pt
\noindent
{\bf (3) Disks of spirals and S0 galaxies.~}
These have absolute magnitudes $-21 < M_B < -17$, and central
surface-brightnesses close to 21.6 $B$ mag arcsec$^{-2}$ (Freeman's law
[11]),
which is fainter than the central surface-brightness of almost any sequence (1)
ellipticals.  At faint magnitudes $M_B > -18$, this sequence blends into the 
dwarf galaxy sequence -- see (4) below.
\vskip 1pt
\noindent {\bf (4) Dwarf Galaxies.~}
These have $M_B > -18$ and form a sequence of fainter central
surface-brightness with decreasing total luminosity.  They are also 
increasingly dark matter dominated at fainter magnitudes [25,16]:
the faintest dwarfs like
Draco and Ursa Minor have mass-to-light ratios $\sim 10^2$ within their
core-fitting radii [1] and total mass-to-light ratios
that are probably much larger than this [28]. 
It is because of their high dark matter content that a link between
the mass functions of these faint dwarfs and the power spectrum of
primordial fluctuations on small scales is suggested.
Dwarf galaxies are typically one of two morphological types: (i) dwarf
irregulars (dIrr) or (ii) dwarf spheroidals (dSph, alternatively called
dwarf ellipticals, or dE).  These types of galaxies have similar scaling
laws (they occupy the same region on the Binggeli diagram), but dwarf 
irregulars have bluer colors and more disturbed morphologies.  
Familiar examples in
the Local Group are the LMC (dIrr), NGC 205 (a bright dSph with 
$M_B \sim -15$), and Draco (a faint low surface-brightness dSph with
$M_B \sim -8$).
\vskip 1pt
\noindent
There do exist other, rare galaxies which do not fit into the classifications
above.  Examples are huge low surface-brightness giants like Malin 1, and blue
compact dwarfs, which appear to have been caught in a short burst of
extreme star formation.  Throughout the rest of this paper, by
``dwarf galaxy'', I will mean sequence (4) objects.

\section{Introduction -- clusters}
Clusters of galaxies have proved a popular environment for determining
the LF.  This is because the distances to the galaxies there are
known (at least in a statistical sense) so that the LF can be determined
by photometry alone and spectroscopic redshifts are not needed.  This in
turn means that the LF can be measured down to very faint limits.

But there is a price to be paid for the cosmologist.  Cluster galaxies are
in dense environments where the crossing time plus age is less than or
comparable to the Hubble time.  These galaxies have therefore had their
properties shaped by cluster-related processes e.g.~ram-pressure stripping
of gas from the galaxies.  Measurements of the galaxy LF in clusters 
might then
tell us more about these cluster-related processes 
than about cosmological parameters like the shape of the power spectrum
on small mass scales.
This is particularly true of very dense environments like the
centers of rich clusters with large fractions of elliptical
galaxies.  It is less true of diffuse spiral-rich environments and the
outer parts of clusters.

In Figure 1, for a sample of clusters with LF determinations, I present the
velociy dispersion (an indication of the mass of the cluster) as a function
of the cluster redshift $z$.  Also shown on this figure are lines 
corresponding to 200 kpc = 10 arcmin -- this is where a rich cluster core
roughly fits into a single 2K CCD frame on a 2 to 4 meter telescope -- and
corresponding to where a typical dSph galaxy with $R=24$ has a scale-length
(as inferred from the Binggeli plot) of 1 arcsec -- to the right of 
this line, most faint cluster 
galaxies detected have intrinsic scale-lengths 
smaller than the seeing and so cannot be readily discriminated from
background galaxies on morphological grounds.

\begin{figure}
 \vspace{300pt}
 \caption{
  The velocity dispersions and redshifts for some clusters with
  recent LF determinations}
 \end{figure}

The techniques used to determine the LF in a particular cluster
are determined by its position on Figure 1.  In simplest terms, clusters
are either on the left of the dashed lines (hereafter I call these
``nearby clusters'') or between or to the right of the dashed
lines (I call these ``distant clusters'').

Most early work in this subject that reached absolute
magnitudes faint enough to
probe dwarf galaxies (those in sequence (4) of the Binggeli diagram)
were photographic measurements of nearby clusters e.g.~Virgo [29]
and Fornax [10].  
Clusters significantly more 
distant than Coma (see e.g.~ref.~35 for a large
photographic survey of Coma)
could not be 
studied in this way because (i) photographic plates
were not sensitive enough to probe the bulk of the dwarf population,
and (ii) cluster members could not be readily distinguished from
background/foreground galaxies.  

However, the advent of 2K CCDs on 2 -- 4 m telescopes allowed very precise
statistical determinations of field galaxy counts down to very faint
magnitudes (e.g.~refs.~43 and 17). 
Background/foreground contamination in cluster fields could then be
estimated and corrected for down to $R=25$, and the LFs of several rich
clusters out to $z = 0.2$ were measured.  
The main limitation of these studies was that the errors at the
faintest magnitudes were quite large because of uncertainty in the
field-to-field variance of the background.  However, large samples of
clusters could be studied, and these errors reduced when combining samples
of similar clusters.  In Section 5, some results of this type will be
presented and discussed.

These 2K CCDs are not well suited to studies of nearby clusters,
although individual dwarfs can easily be identified and no background
subtraction is needed.  This is because the CCDs are small and each
exposure can only cover a negligible fraction of the cluster core
(we are to the left of the 200 kpc = 10 arcmin line in Figure 1).  Counting
statistics are then crippling -- a typical 2K CCD field in the center of
the Virgo Cluster will have no or at most 1 or 2 dwarfs brighter than
$R=22$.

More recently, 8K and larger CCDs have become available on 4 m telescopes
(e.g.~the UH 8K mosaic on the 3.6 m CFHT [23]).
Large samples of galaxies in nearby clusters can now be obtained and
the LFs measured accurately down to very faint magnitudes.
Because no background subtraction is required, the only errors at faint
magnitudes come from counting statistics -- the errors are consequently
much smaller than the errors at the faint end for the distant clusters,
at least for reasonably dense nearby clusters like Virgo and Fornax.
An alternative approach has been employed by Phillipps et al.~[27] who
use sensitive median-stacked UK Schmidt plates [32] 
to measure the Virgo cluster LF down to $M_R \sim -11.5$
(see the paper by Jones et al.~in these proceedings -- this was one of
the most significant new results at this meeting).
A project in collaboration with B.~Tully (University of Hawaii) has been
undertaken to measure the Ursa Major LF but this cluster is sufficiently
diffuse that counting statistics are still a significant problem.

\section{Techniques}
For the distant clusters, a number of specialized techniques have evolved
[9,4,38] that
optimize the accuracy to which the LF can be measured at faint magnitudes.
A recurrent theme is the importance of characterizing the
background/foreground contamination and its uncertainty, as this is the
main source of the error in the faintest points of the LF.  The
contamination is severe: the background/foreground galaxy counts are
comparable to or greater than the cluster counts at $R > 19$, even in
the centers of rich clusters like Coma. 
For example, high photometric accuracy in the zero-point needs to be
achieved because the background counts are a strongly-varying function 
of magnitude and we must be sure that we are subtracting
exactly the right amount of background galaxies at each magnitude.
Other systematic sources of error (e.g.~Galactic extinction, possible
extinction from dust in the clusters, stellar contamination,
globular cluster contamination, giant galaxy halo substructure, distortion
of the background counts by gravitational lensing by the cluster dark
matter) are important too and all need to be considered.  The reader
is referred to the papers listed above for details.

Measurements of nearby clusters as outlined in Section 4 pose different
problems.  Background subtraction is no longer a major
problem because the faintest dwarfs have scale-lengths that are
significantly larger than the seeing so that cluster members can be
identified by their morphologies.  Therefore the LF of Virgo measured by
Phillipps et al.~[27] is very well constrained at the faint end.  
Measurements in diffuse groups like Ursa Major are more difficult because
counting statistics there are poor.  Also, the surface density of
low-surface-brightness (LSB) field galaxies which look similar to
cluster dwarfs needs to be well characterized (for Virgo this is not a
serious problem because the cluster dwarfs outnumber background LSB
galaxies by a large factor in this relatively dense cluster -- see
the paper by Jones et al.~in these proceedings).

\section{The luminosity functions of the cores of rich clusters}
In Figure 2, I present four LFs from my earlier work [38 -- 41]. 
Other published LFs (in addition to the ones already listed)
are presented in references 12, 19, 20, 33, 34, and 47.

\begin{figure}
 \vspace{300pt}
 \caption{
The $R$-band luminosity functions of A 262 ($z=0.016$), Coma
($z=0.023$), A 1795 ($z=0.063$), and A 665
($z=0.182$).  In converting apparent to absolute magnitudes,
I assume $H_0$ = 75 km s$^{-1}$ Mpc$^{-1}$ and $\Omega_0 = 1$.  
 }
 \end{figure}

For A 262 (a poor cluster at $z = 0.016$), signs of a turn-up at the
faint-end ($M_R < -14$) are seen, but the LF is not well-constrained at
bright magnitudes here because the galaxy density is too low.  In Coma,
a richer cluster at approximately the same distance as A 262, a similar
turn-up at the faint-end is seen, but a far more shallow LF is observed
at brighter magnitudes.  In both cases, the statistics are poor at the
faint-end because the field-to-field variance of the background is so
large -- nevertheless, in both cases the turn-up at the faint-end is
statistically significant and the LF is not fit by either a power-law
or a Schechter function [30] over the entire range $-20 < M_R < -11$).
For the more distant clusters, A 1795 and A 665, a flatter LF is seen
but the very faint magnitudes at which the turn-up is seen in A 262 and
Coma are not probed in these clusters.

The approximate similarity of the form of the LFs in the four clusters in
Figure 2 suggests that it might be productive to combine the individual cluster
LFs in order to improve the statistics.  This is done in Figure 3.
When performing this calculation, I found that each individual cluster
LF is consistent with the composite LF to a high degree of statistical
significance.  This is suggestive that the LF of galaxies in the cores
of rich clusters is remarkably constant.  Also, 
the Fornax LF of Ferguson [10] is also consistent with this composite
function at a high level of significance down to $M_B = -13$
(see ref.~42 -- this is
particularly intriguing because the errors in the Fornax sample at the
faint end come only from counting statistics and are quite small).

\begin{figure}
 \vspace{330pt}
 \caption{
Composite $R$-band luminosity function of
rich clusters (derived using the LFs from Coma, A 2199, A 1795,
A 1146, A 665, and A 963).  This represents the weighted average of
the individual LFs, normalized to a projected galaxy density corresponding
to that of a typical Abell richness 2 cluster.
Local values of $\alpha (M_R)$ derived by measuring the slope between
$M_R - 1$ and $M_R + 1$, and their uncertainties, are given too.
Slopes corresponding to $\alpha = -1$ and $\alpha = -2$ are also shown.
 } 
 \end{figure}

The composite LF in Figure 3 has much better statistics than the LFs 
shown in Figure 2, now that a sample of clusters have been combined.
The composite LF shows the following properties:
\vskip 1pt
\noindent
1) the LF shows significant curvature at both the bright and faint-ends.
Neither a power-law nor Schechter function [30] provides a good
fit to the data.  
A similar conclusion was made for the Coma LF in Figure 2, but the
smaller error bars in Figure 3 now make this true at a far higher
level of confidence.
\vskip 1pt
\noindent
2) at no magnitude is $\alpha = -1$ a good fit to the data.
We note however the peculiar result that the LF is flattest
($\alpha = -1.2$) at about $M_R = -19$, which is the transition magnitude 
between giants ellipticals and dwarfs in the Binggeli diagram (most giant
galaxies in the centers of rich clusters are ellipticals).
This is suggestive of a conspiracy in which the giant galaxy LF is falling
by an amount almost exactly compensated for by the rise in the dwarf
galaxy LF at $M_R = -19$).
\vskip 1pt
\noindent
3) the turn-up at the faint-end appears to be very steep.  The new
results for Virgo (ref.~27,
Jones et al.~these proceedings) suggest a value
of $\alpha = -2.1$ at $M_R = -12$.  The statistics that Phillipps, Jones
and collaborators got for Virgo are significantly better than I show
for A 262 or Coma in Figure 2, because in Virgo a background subtraction
is not required.  

On theoretical grounds, this steep LF is not predicted to extend to
indefinitely faint magnitudes because we expect photoionization from the
UV background to suppress the conversion of gas into stars in very low mass
systems [36].  On observational grounds, we arrive
at the same prediction from the requirement that the integrated light from
small galaxies not overproduce the measured intracluster light 
(see Section 8 below).  It is, however, interesting to note that between
about $M_R = -13$ and whenever this turnover occurs, the LF has approximately
the same shape ($\alpha \sim -2$)
as the mass function predicted [49] from Press-Schechter
theory (i.e.~if Press-Schechter theory is valid in this
context, then over this magnitude
range, the combined efficiency of all the baryonic processes which  
convert gas to stars is independent of the mass of the galaxy).

An important caveat is that the LFs presented in Figures 2 and 3 are
only valid for the cores of rich clusters, which are dense elliptical-rich
environments.  Perhaps the agreement between the composite LF and the
Fornax LF outlined above suggests that it is the elliptical galaxy fraction,
not the total galaxy density, that is important (Fornax is anomalously
elliptical-rich).  There is good evidence [8] 
that the ellipticals evolve differently from all other giant galaxies
(including S0s) in clusters, so it is perhaps unexpected if the 
global form of the LF is the same in elliptical-rich and elliptical-poor
environments.   

\section{Comparison between clusters and the field}
The largest sample of field galaxies to date is the Las Campanas Redshift
Survey (LCRS; ref.~18), which has $1.9 \times 10^4$ galaxies.
For $-18 < M_R < -16$ (approximately
$-16.5 < M_B < -14.5$), the LCRS has a slope of $\alpha = -1.39 \pm 0.11
$.  The cluster slope over
the same magnitude range is $\alpha = -1.44 \pm 0.13
$.

There are, however, differences.  Most of the faintest 
galaxies seen in the LCRS are emission-line galaxies
where as most of the faintest galaxies in clusters are 
gas-poor dSph galaxies.
Therefore, even though the LF slope is similar in the two environments
for $-18 < M_R < -16$, the LF is dominated by the contributions from
different types of galaxies in each. 

Also, the field and cluster LFs are
probably very different much fainter than
$M_R = -14$.  The steep turn-up seen in Virgo [27] for $M_R > -14$ certainly
does not happen in the Local Group [45], for example.  Although the Local Group
may have several undiscovered low surface-brightness members at the very
faintest magnitudes like $M_R \sim -9$ (the magnitude of Draco and
Ursa Minor), it probably
does not have many brighter than $M_R = -12$.
Therefore incompleteness is not the reason for the discrepancy in the
LFs here.

\section{The types of the faintest galaxies in rich clusters}
In the previous two sections, I have presented evidence that supports
the presence of a large number of faint galaxies in clusters, but
have not shown what type these galaxies are.  The scale-lengths are
consistent with either a dSph or dIrr interpretation (recall from Section
2 that these galaxies occupy the same part of the Binggeli diagram).  They
are not consistent with the faintest galaxies being low luminosity
compact ellipticals like M32.
These faint galaxies have scale-lengths comparable to
or smaller than the seeing, so that their
detailed morphologies cannot
be used to tell whether they are dSphs or dIrrs.

The best discriminant between these galaxy types in the
absence of spectroscopic information is the color of the galaxies.
In all the distant clusters shown on Figure 1, I found that the colors
of the faintest galaxies were consistent with them being dSphs and not
dIrrs (the K corrections are given in ref.~41).  
Other authors (see the list of references in Section 5) conclude
similarly.  
In Virgo, the galaxies are sufficiently nearby that morphological 
information can be used -- both Sandage et al.~[29] and Phillipps
et al.~[27] find that the faintest galaxies that they detect are dSphs
as well.

\section{The lowest surface-brightness galaxies}
With measurements of this type there is always the long-standing worry
that we are missing many LSB galaxies and so are underestimating the
cluster galaxy counts at each magnitude [7].  Then the LF that
we derive could be seriously in error.

In all clusters I studied (the distant clusters in Figure 1), I found that
on going to fainter magnitudes, I did not find systematically more galaxies
that had surface-brightnesses close to my detection threshold.  This is
indicative that I am not missing many galaxies because they have
surface-brightnesses too low to be detected, but is far from being a
rigorous statement to this effect.  There may be, for example,
an entire population
of extremely LSB galaxies with surface-brightnesses far lower than my
detection threshold.  A more formal analysis is suggested.

So let us define (as Phillipps and Disney [26] did for spiral galaxies)  
a bivariate surface-brightness / luminosity distribution function
in the usual way: $\phi (\mu,L) {\rm d} \mu {\rm d} L$ is the number
density of galaxies with surface-brightnesses between $\mu$ and
$\mu + {\rm d} \mu$ and luminosities between $L$ and $L + {\rm d} L$.
When we measure a LF (as in the rest of this paper), what we are
really measuring is the contraction of this 
bivariate function over high 
surface-brightnesses: $\phi (L)_{\rm measured} = \int_{\mu_c}^{\infty}
\phi (\mu,L) {\rm d} \mu $ where $\mu_c$ is the detection
threshold.

There is a constraint on the other side of this contraction (i.e.~over
low surface-brightnesses): $\int_{0}^{\infty} \int_{0}^{\mu_c} L 
\phi (\mu,L) {\rm d} \mu {\rm d} L$ must not exceed the intracluster
background light (e.g.~refs.~22 and 37, 
and more recently refs.~31, 44, and 46). 
This is only a limit because 
much of the intracluster background light might come from stars that 
have been tidally stripped from the giant galaxies; such stars are
not associated with LSB galaxies.
This constraint is however severe enough to rule out extremely steep LFs
in clusters (e.g.~the steep ($\alpha \sim -2.8$) LF recently
proposed by Loveday [21] for the field
would not work in the Coma cluster because of this constraint).

\section{The mass function}
Everything presented so far in this review
has dealt with the luminosity function of
galaxies, which
is not the same as the mass function (by mass function here I am referring
to total mass, where the dark matter content of individual galaxies is
included).
The two, however, probably are 
fairly similar, unless there exist a significant dispersion in the
mass-to-light vs.~light distribution function for galaxies in clusters.
Clusters of galaxies possess a great deal of dark matter so that it is not
impossible that they contain many galaxies that are almost completely
dark.
The mass function would be a useful probe of the fluctuation spectrum,
since it bypasses all the complex
physics of the baryonic component outlined in
Section 1 (in the cores of rich
clusters, however, there remains the caveat that the
dark-matter properties of galaxies might be significantly affected by
cluster-related processes e.g.~tidal stripping).

The mass function is not straightforward to measure.  The best hope is
to determine it through gravitational lensing measurements, for example
(i) galaxy-galaxy strong lensing, or (ii) the Natarajan-Kneib [24] 
method, which attributes granularity in the weak lensing
shear map to lensing by
individual galaxy halos.  In the upcoming age of 8 and 10 m telescopes, it
is likely that these kinds of measurements will provide important 
constraints. 


%
\vfill
\end{document}